# On the anti-cancer effect of cold atmospheric plasma and the possible role of catalase-dependent apoptotic pathways

**Charlotta Bengtson * and Annemie Bogaerts**

Research Group PLASMANT, Department of Chemistry, University of Antwerp, Universiteitsplein 1, B-2610 Wilrijk-Antwerp, Belgium
* Correspondence: charlotta.bengtson@uantwerpen.be

**Abstract:** Cold atmospheric plasma (CAP) is a promising new agent for (selective) cancer treatment, but the underlying cause of the anti-cancer effect of CAP is not well understood yet. Among different theories and observations, one theory in particular has been postulated in great detail and consists of a very complex network of reactions that are claimed to account for the anti-cancer effect of CAP. Here, the key concept is a reactivation of two specific apoptotic cell signaling pathways through catalase inactivation caused by CAP. Thus, it is postulated that the anti-cancer effect of CAP is due to its ability to inactivate catalase, either directly or indirectly. A theoretical investigation of the proposed theory, especially the role of catalase inactivation, can contribute to the understanding of the underlying cause of the anti-cancer effect of CAP. In the present study, we develop a mathematical model to analyze the proposed catalase-dependent anti-cancer effect of CAP. Our results show that a catalase-dependent reactivation of the two apoptotic pathways of interest is unlikely to contribute to the observed anti-cancer effect of CAP. Thus, we believe that other theories of the underlying cause should be considered and evaluated to gain knowledge about the principles of CAP-induced cancer cell death.

## 1. Introduction

Recently, the application of cold atmospheric plasma (CAP) on cancer tissue has emerged as a novel, promising cancer treatment, and up to now, CAP has shown a significant effect in over 20 different types of cancer cell lines (*in vitro*) including brain cancer [1,2], skin cancer [3–5], breast cancer [6,7], colorectal cancer [8,9], lung cancer [10,11], cervical cancer [12,13], and leukemia [14,15]. The first results from the clinical use of CAP in cancer treatment are equally encouraging [16–18]. A special feature - and great advantage compared to most conventional cancer treatments - is the possibility to selectively eliminate cancer cells while leaving normal cells unaffected [8,10,19–28], although this selectivity is not always observed and depends on the treatment conditions [29].

It is generally accepted that the anti-cancer effect of CAP, to a great extent, is associated with reactive oxygen species (ROS) and reactive nitrogen species (RNS) generated in CAP. Especially, a rise of intracellular ROS in cancer cells (but not in normal cells) upon CAP treatment has been reported [22,24,30–32]. The significance of this intracellular rise of ROS has been experimentally verified by the observation that CAP treatment fails to eliminate cancer cells when these are pretreated with intracellular ROS scavengers [2,30,31]. Furthermore, CAP has been shown to induce apoptosis in cancer cells in vitro and *in vivo* [10,33–42]. In the light of these findings, it has been argued that the underlying cause of the anti-cancer effect of CAP is related to apoptosis-induction mediated by ROS and RNS [30,39,41,43–46] and several different theories of such an underlying (selective) cause of ROS- and RNS-mediated apoptosis-induction have been proposed. These include the following theories:

- Higher influx of hydrogen peroxide due to a higher density of aquaporins in the cell membrane of cancer cells [47].
- Immunogenic cell death triggered by CAP [48–52].

- Membrane pore formation and subsequent influx of (CAP-generated) ROS and RNS in cancer cells [53,54].

In addition, a theory that postulates a very complex and detailed network of reactions, which is claimed to account for the underlying cause of the anti-cancer effect of CAP, has been frequently published; see e.g., refs. [37,55–60]. This theory introduces a whole new concept of the cause of the anti-cancer effect of CAP; instead of the common hypothesis that CAP-generated ROS and RNS are directly responsible for cancer cell apoptosis, it is claimed that CAP merely acts as a trigger to (re)activate a selective self-destruction process specific for cancer cells. More explicitly, the underlying cause, according to this theory, is originating from two different autocrine, apoptotic, signaling pathways occurring in the extracellular compartment of cancer cells. It also involves the concept of membrane-associated catalase, which will inhibit these two signaling pathways from inducing apoptosis. Ultimately, CAP will reactivate these pathways through inactivation of the protective membrane-associated catalase. Singlet oxygen, either generated directly in the CAP source, or indirectly by ROS and RNS in CAP, has been proposed as the species causing catalase inactivation, see e.g., refs. [35,55–61]. The end product of both signaling pathways is the hydroxyl radical, which causes apoptosis-induction through lipid peroxidation in the cell membrane.

So far, the introduced theory has only been justified by experimental observations. In addition, the experimental results have been reported in terms of the proportion of apoptotic cells for different experimental conditions, such as whether the extracellular catalase is inactivated or not. Although experiments with addition of different species has been carried out in a systematic manner to prove the existence of the proposed signaling pathways, the underlying mechanisms from catalase inactivation to cell apoptosis is still to a great extent a "black box". Indeed, no quantitative data in terms of concentration of different species exists. Mathematical modeling of the kinetics of the proposed signaling pathways, which consists of a set of chemical reactions, may be used as a complement to try to unveil whether the reactivation of the two apoptotic pathways could account for the mechanisms in the "black box". In other similar systems, a theoretical investigation through mathematical modeling has proven to be a useful approach to increase the knowledge of the mechanisms of cell antioxidant defense and cell signaling, see e.g., refs. [62–74]. An advantage of a mathematical model is that it allows one to probe the system's behavior in ways that would not be possible in experiments.

In the present study, we theoretically investigate the catalase-dependent apoptotic pathways introduced in refs. [37,55–60]. By mathematical modeling of the reaction kinetics of these pathways, we analyze and evaluate this theory of the underlying cause of the anti-cancer effect of CAP from a kinetic point of view. In particular, we explore the catalase-dependency of the pathways by quantifying the amount of generated hydroxyl radicals as a function of catalase concentration. Due to the lack of experimental parameter values, we use mathematical logic to reject certain scenario's (corresponding to implausible outcomes) by analyzing the kinetics at optimizing conditions and conditions corresponding to physical limits of the system. The aim of this study is to contribute to knowledge and understanding of the underlying cause of the (selective) anti-cancer effect of CAP, knowledge that could be useful in the process of optimizing the treatment conditions of CAP for clinical applications. We find that the theory proposed in refs. [37,55–60] unlikely accounts for the observed anti-cancer effect of CAP and should be revised in order to capture the underlying cause sufficiently well to understand the connection between CAP treatment conditions and cancer cell death.

As a framework for the modeling, we first briefly introduce the theory of the catalase-dependent apoptotic pathways and how they may be reactivated by CAP.

## 2. Anti-cancer effect of CAP by reactivation of catalase-dependent apoptotic pathways

In the theory presented in refs. [37,55–60], cancer cells are characterized by two important features:

- Generation of extracellular superoxide anions.
- Membrane-associated catalase.

The extracellular superoxide anion-generating enzyme NOX1 has been connected to cancer cell proliferation (i.e., the growth of cancer cells) [75–86], and the generation of superoxide anions by several cancer cell lines has been reported [87,88]. Furthermore, there are studies showing that cancer progression requires the expression of membrane-associated catalase [89–93]. How these two features are believed to cause a selective anti-cancer effect of CAP, will be explained in the following three subsections.

## 2.1. Apoptosis-inducing signaling pathways originating from superoxide anions

According to the theory in refs. [37,55–60], the extracellular NOX1-generated superoxide anions are also the precursor of two apoptosis-inducing signaling pathways; the hypochlorous acid pathway and the nitric oxide/peroxynitrite pathway [89,90,94–96]. Together these pathways form what will be referred to as a reaction network. Both pathways in the reaction network result in the generation of hydroxyl radicals, which cause apoptosis induction through lipid peroxidation in the cell membrane (cf. above). In the hypochlorous acid pathway, hydrogen peroxide is formed from superoxide anions in a reaction catalyzed by superoxide dismutase (SOD). Subsequently, hydrogen peroxide is used to synthesize hypochlorous acid in a reaction catalyzed by peroxidase (POD). Thereafter, hypochlorous acid reacts with superoxide anions to form hydroxyl radicals. In the nitric oxide/peroxynitrite pathway, superoxide anions first react with nitric oxide to form peroxynitrite. Subsequently, protonation of peroxynitrite forms peroxynitrous acid, which decomposes into nitrogen dioxide and hydroxyl radicals. For a schematic illustration of both pathways in the reaction network, see Figure 1.

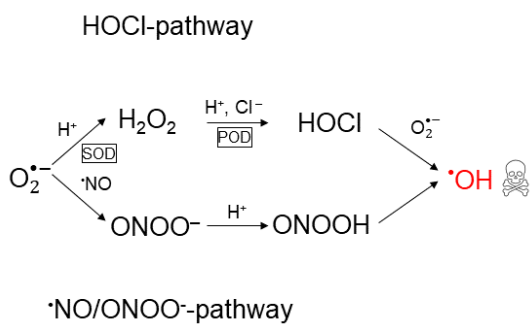

**Figure 1.** Schematic illustration of the reaction network of apoptosis-inducing signaling pathways: the hypochlorous acid- and the nitric oxide/peroxynitrite pathway. Both pathways originate from superoxide anions and result in the formation of hydroxyl radicals, which are responsible for apoptosis-induction.

Note that although the ROS's and RNS's involved in the reaction network also may be generated by reactions in the CAP source, see e.g., [97,98], the theory in refs. [37,55–60] only accounts for apoptosis-induction by the hydroxyl radicals generated from the hypochlorous acid- and nitric oxide/peroxynitrite pathways. Neither does the theory account for any other possible physical effect, such as microwave and UV emission, that may be of importance in the context of the anti-cancer effect of CAP.

## 2.2. The protective role of catalase

According to the theory in refs. [37,55–60], the function of membrane-associated catalase in the extracellular compartment of cancer cells is to maintain the concentration of generated hydroxyl radicals below the threshold of apoptosis-induction. Catalase decomposes hydrogen peroxide into water and oxygen and thus removes the substrate for hypochlorous acid production [91,93]. In addition, catalase can reduce the formation of peroxynitrite (through oxidation of nitric oxide [99]) as well as decompose peroxynitrite into nitrite and oxygen [34,93,100]. Since the formation of peroxynitrite through reaction of nitric oxide and superoxide anions is very fast (much faster than the interaction between nitric oxide and catalase), it is reasonable to assume that the latter constitutes the major contribution to the peroxynitrite lowering effect of catalase. This conclusion was also drawn in e.g., [34]. For a schematic illustration of the effect of catalase on the reaction network presented in Figure 1, see Figure 2.

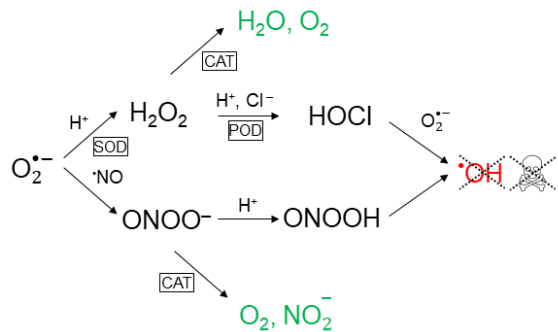

**Figure 2.** Catalase prevents the formation of hydroxyl radicals in both apoptosis-inducing signaling pathways. In the hypochlorous acid pathway, this occurs through decomposition of hydrogen peroxide into water and oxygen, and in the nitric oxide/peroxynitrite pathway, it proceeds through decomposition of peroxynitrite into nitrite and oxygen.

### 2.3. Effect of CAP on cancer cells

According to refs. [37,55–60], the (selective) anti-cancer effect of CAP is due to its effect on catalase; it has been shown that singlet oxygen, which is one of the ROS known to be generated in CAP, has the potential to inactivate antioxidant enzymes, like catalase [101,102]. Thus, CAP may be used as a source of external so-called "primary" singlet oxygen to inactivate the protective membrane-associated catalase in cancer cells. Applying CAP on tissue containing cancer cells will hence cause selective elimination of cancer cells through reactivation of hydroxyl radical generation from the hypochlorous acid- and the nitric oxide/peroxynitrite pathway in the extracellular compartment of these cells. A simplified cartoon picture of the total chain of events presented in Sections 2.1–2.3, can be seen in Figure 3.

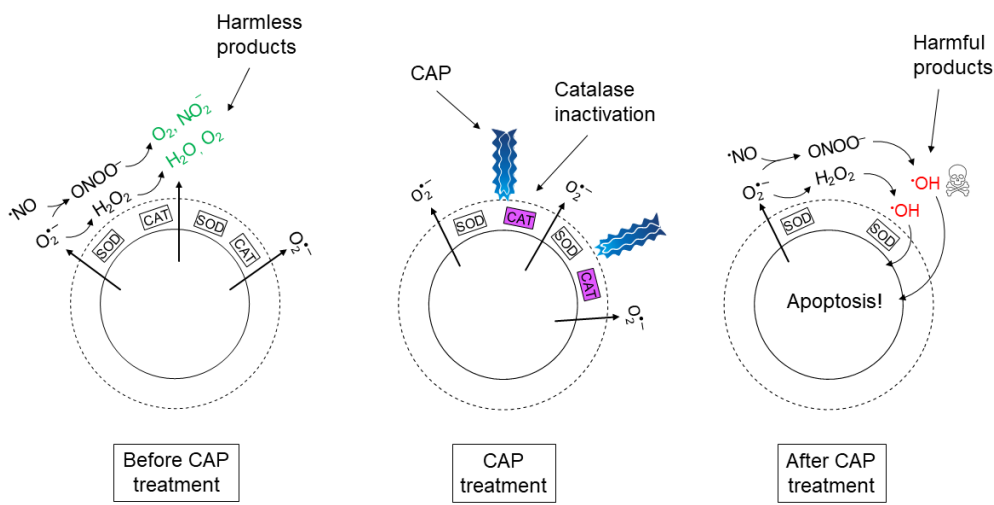

**Figure 3.** Proposed set of events underlying the mechanism of (selective) cancer treatment with cold atmospheric plasma (CAP). The cartoon picture represents a cancer cell and its extracellular compartment before, during, and after CAP treatment.

Another important aspect of the theory of the underlying cause of the anti-cancer effect of CAP, as presented in refs. [37,55–60], is the hypothesis that the non-decomposed hydrogen peroxide and peroxynitrite in the vicinity of inactivated catalase, may generate a burst of so-called "secondary" singlet oxygen, which is propagated to adjacent cancer cells. In this catalase-dependent self-perpetuation of singlet oxygen, the effect of CAP is also predicted to go beyond the surface of the cancer tumor. As an additional contribution to the investigation and discussion of this theory, we also briefly evaluate the plausibility of the hypothesis of catalase inactivation by primary singlet oxygen, as well as by the generation of secondary singlet oxygen (see Section 5.3).

## 3. The mathematical model

In this section, we present the construction and implementation of the mathematical model. We also point out the assumptions and simplifications made in the model development and present the justification of those.

### 3.1. Construction of the mathematical model

Since this is the first attempt to construct a mathematical model of the reaction network of the hypochlorous acid- and nitric oxide/peroxynitrite pathways and its interaction with catalase, as our first simplification, we (initially) choose to restrict the model to include the closed system of the reaction network (catalase-interaction included) itself only. Thus, any other possible interfering pathway or interaction with the surrounding is neglected. We admit that this model will not capture the full complexity of the dynamics of the reaction network in its *in vitro* or *in vivo* context, but it will provide fundamental insights about the kinetics of the reaction network itself and necessary conditions for catalase-dependent reactivation of hydroxyl radical generation. Especially, neglecting interfering pathways will allow us to reject a negative result since the conditions for hydroxyl radical generation is optimal. Indeed, additional pathways would most certainly decrease the generation of hydroxyl radicals from the hypochlorous acid- and nitric oxide/peroxynitrite pathways.

As a second simplification, we assume that the extracellular compartment of a cancer cell can be thought of as a well-mixed reaction vessel. All involved species are thus homogenously distributed. The motivation for only considering the reaction kinetics is as follows: When the spatial dynamics is included in a mathematical model of a reaction network, the rates of reactions of the different species in the network are competing with their diffusional loss. Since the important species in the reaction network all are located close to the cell membrane in the extracellular compartment, we can assume that the local concentration of especially enzymes is high and reactions should forestall diffusion. Thus, diffusion becomes less important since the species most probably will react before diffusing from the site of production. This assumption is justified by a number of studies for the two key species, i.e., nitric oxide and superoxide anions [64,65,103,104]. In these studies, the concentration of nitric oxide, superoxide anions, and related species in the extracellular compartment of activated macrophages was investigated by mathematical modeling. Like cancer cells, macrophage cells generate both nitric oxide and superoxide anions into the extracellular compartment. In ref. [65], it was estimated that superoxide anions are depleted within 1 $\mu m$ above the cell surface and in [64] it was found that both superoxide anions and peroxynitrite only exist in a very thin layer above the cell surface. In all these studies, nitric oxide reached the same steady-state concentration in the vicinity of the cells (despite different conditions such as cell density). Thus, in summary, we are modeling the kinetics of the reaction network in a thin layer above the surface of *one* cancer cell as it would occur in a well-mixed reaction vessel.

Explicitly, we use the mathematical model to analyze the behavior of the dependent variable $y$ defined as

$$y(\overline{x}) = [^\bullet OH]_{max}^{\overline{x}}.$$

Here, $\overline{x}$ denotes the set of independent variables that are varied in a particular system, which in our case is $[CAT]_0$, i.e., $\overline{x} = [CAT]_0$ and $[CAT]_0$ denotes the initial concentration of catalase, while $[^\bullet OH]_{max}^{\overline{x}}$ is the temporal maximum of the concentration of hydroxyl radicals. Thus, $[^\bullet OH]_{max}^{\overline{x}}$ is the dependent variables in our calculations and it will capture the peak value of the hydroxyl radical concentration for a certain initial catalase concentration.

We assume that the cellular generation rates of nitric oxide and superoxide anions exactly balance the rates of consumption before $t = 0$, i.e., the concentrations of nitric oxide and superoxide anions are constant (steady-state) before the perturbation of the kinetics by the introduction of catalase in the reaction network. This perturbation is assumed to occur instantly at $t = 0$ and the effect is monitored thereafter. Thus, regarding the extracellular compartment as a reaction vessel in an experiment, the time $t = 0$ corresponds to the moment when the reaction vessel is prepared, and the reading starts.

Ideally, the model should be compared and, if necessary, calibrated to experimental results, but to the best of our knowledge, there are no experimental studies for the catalase-dependency of the concentration of generated hydroxyl radicals in the extracellular compartment of cancer cells. Neither are there studies where an explicit correlation between the amount of extracellularly generated hydroxyl radicals and cancer cell apoptosis is reported - information that ultimately would be required as a reference value of when the hydroxyl radicals production can be considered reactivated. For externally added hydroxyl radical, the critical concentration to cause apoptosis-induction for lung cancer H460 cells, is around $0.3 \times 10^{16}\ cm^{-3}$ [105,106]. This corresponds to $[^\bullet OH] = 5 \times 10^{-6}\ M$, which could be a value to refer back to when we evaluate our results.

### 3.2. Reaction network

The reaction network, representing both apoptotic pathways in the extracellular compartment of a cancer cell, and their interaction with catalase, consists of the following reactions:

$$2O_2^{\bullet -} + 2H^+ \xrightarrow[SOD]{k_1} H_2O_2 + O_2, \tag{1}$$

$$^\bullet NO + O_2^{\bullet -} \underset{k_{-2}}{\overset{k_2}{\rightleftharpoons}} ONOO^-, \tag{2}$$

$$2H_2O_2 \xrightarrow[CAT]{k_3} O_2 + 2H_2O, \tag{3}$$

$$2ONOO^- \xrightarrow[CAT]{k_4} O_2 + 2NO_2^-, \tag{4}$$

$$H_2O_2 + Cl^- + H^+ \xrightarrow[POD]{k_5} HOCl + H_2O, \tag{5}$$

$$HOCl + O_2^{\bullet -} \underset{k_{-6}}{\overset{k_6}{\rightleftharpoons}} {^\bullet OH} + Cl^- + O_2, \tag{6}$$

$$ONOO^- + H^+ \underset{k_{-7}}{\overset{k_7}{\rightleftharpoons}} ONOOH, \tag{7}$$

$$ONOOH \underset{k_{-8}}{\overset{k_8}{\rightleftharpoons}} {^\bullet NO_2} + {^\bullet OH}, \tag{8}$$

where $k_i$ is the rate constant of the forward reaction for reaction $(i)$ and $k_{-i}$ is the rate constant for the backward reaction. Enzyme-catalyzed reactions are assumed to only proceed in the forward direction, in accordance with the assumptions for which experimental parameter values exist [107,108].

Reaction (1), (3), (5) and (6) form the hypochlorous acid pathway, whereas reaction (2), (4), (7) and (8) form the nitric oxide/peroxynitrite pathway.

### 3.3. Rate equations

The kinetics of a reaction network is given by the set of rate equations describing the rate of production and consumption of each species in the network. Here, we summarize the explicit equations, as well as the assumptions we are making. A more detailed description of the rate equations and the references from where they have been taken, can be found in the Appendix A.1.

Note that the reactions (1)–(8) are not elementary reactions, but rather consist of many different elementary reactions that we do not know about. The information available is the rate constants and reaction order for each species in the overall reaction, as they appear in experiments [100,107–118]. Hence, not all rate equations follow the law of mass action, but have other, experimentally observed, forms. More information is given in the Appendix A.1.

We assume that $[H^+]$ is kept constant over time (due to a constant generation by proton pumps), i.e.,

$$\frac{d[H^+]}{dt} = 0.$$

Furthermore, we assume that $[Cl^-]$ is kept constant over time (due to a high physiological concentration compared to the other species), i.e.,

$$\frac{d[Cl^-]}{dt} = 0.$$

Finally, we assume that the catalysts, $[SOD], [CAT]$ and $[POD]$, are kept constant over time, i.e.,

$$\frac{d[E]}{dt} = 0,$$

where $E$ denotes $SOD, CAT$ or $POD$.

The set of coupled rate equations, solving the time-dependence of the concentration of the different species, is then given by (see Appendix A.1 for more details):

$$\frac{d[O_2^{\bullet-}]}{dt} = -[O_2^{\bullet-}](k_1[SOD] + k_2[^{\bullet}NO] + k_6[HOCl]) + k_{-2}[ONOO^-]$$
$$+ k_{-6}[^{\bullet}OH][Cl^-][O_2],$$

$$\frac{d[H^+]}{dt} = 0,$$

$$\frac{d[H_2O_2]}{dt} = \frac{1}{2}k_1[O_2^{\bullet-}][SOD]$$
$$- [H_2O_2]\left(2k_3[CAT]_0\frac{1}{K_3 + [H_2O_2]} + k_5[POD]_0\frac{1}{K_5 + [H_2O_2]}\right),$$

$$\frac{d[^{\bullet}NO]}{dt} = -k_2[^{\bullet}NO][O_2^{\bullet-}] + k_{-2}[ONOO^-],$$

$$\frac{d[ONOO^-]}{dt} = k_2[^{\bullet}NO][O_2^{\bullet-}] - [ONOO^-](k_{-2} + k_4[CAT] + k_7[H^+]) + k_{-7}[ONOOH],$$

$$\frac{d[Cl^-]}{dt} = 0,$$

$$\frac{d[HOCl]}{dt} = k_5[POD]_0\frac{[H_2O_2]}{K_5 + [H_2O_2]} - k_6[HOCl][O_2^{\bullet-}] + k_{-6}[^{\bullet}OH][Cl^-][O_2],$$

$$\frac{d[^{\bullet}OH]}{dt} = k_6[HOCl][O_2^{\bullet-}] - [^{\bullet}OH](k_{-6}[Cl^-][O_2] + k_{-8}[^{\bullet}NO_2]) + k_8[ONOOH],$$

$$\frac{d[ONOOH]}{dt} = k_7[ONOO^-][H^+] - [ONOOH](k_{-7} + k_8) + k_{-8}[^{\bullet}NO_2][^{\bullet}OH],$$

$$\frac{d[^{\bullet}NO_2]}{dt} = k_8[ONOOH] - k_{-8}[^{\bullet}NO_2][^{\bullet}OH],$$

$$\frac{d[SOD]}{dt} = 0,$$

$$\frac{d[POD]}{dt} = 0,$$

$$\frac{d[CAT]}{dt} = 0.$$

Here, $K_i$ is the Michaelis-Menten constant for reaction $(i)$.

Note that even though the set of rate equations are solved with respect to $t$, we are in this study not interested in the explicit time-dependence of each species but only the maximal concentration of hydroxyl radicals as a function of the catalase concentration (see Section 3.1).

## 4. Numerical details

In this section, we provide details of the numerical calculations, especially the parameter values that we use in the calculations.

### 4.1. Parameter values

The rate constants that we use in this study are collected from the literature, and in the case several different values are reported, we use the value that optimizes the production of hydroxyl radicals to create the upper limit for the hydroxyl radical production (see the Appendix A.1, for more details).

The rate constants for the non-enzyme catalyzed reactions, i.e., reactions $(2), (6), (7)$ and $(8)$, are summarized in Table 1. If not explicitly stated otherwise, the rate constants are for $T = 37\ °C$ and $pH \sim 7$.

**Table 1.** Parameter values of the rate constants of the non-enzyme catalyzed reactions.

| Rate Constant | Value | Reference | Remark |
|---|---|---|---|
| $k_2$ | $1.7 \times 10^{10}\ M^{-1}s^{-1}$ | [109] | No information about $pH$. |
| $k_{-2}$ | $17 \times 10^{-3}\ s^{-1}$ | [116,117] | $T = 20 - 25°C$ |
| $k_6$ | $7.5 \times 10^6\ M^{-1}s^{-1}$ | [110] | $T \sim 20°C, pH = 5.5$ |
| $k_{-6}$ | 0 | | Assigned |
| $k_7$ | $10^{10}\ M^{-1}s^{-1}$ | [112] | |
| $k_{-7}$ | $k_7 K_a, pK_a = 6.8$ | [118] | |
| $k_8$ | $0.6\ s^{-1}$ | [114] | $T = 25\ °C$, no information about $pH$. |
| $k_{-8}$ | $3.5 \times 10^9\ M^{-1}s^{-1}$ | [115] | $pH = 9.5$, no information about $T$. |

The kinetic parameter values of the enzyme catalyzed reactions, i.e., reactions $(1), (3), (4)$ and $(5)$, are summarized in Table 2. If not explicitly stated otherwise, the parameter values are for *homo sapiens* and for $T = 37\ °C$ and $pH \sim 7$.

**Table 2.** Parameter values of the rate constants and Michaelis-Menten constants of the enzyme catalyzed reactions.

| Enzyme | Kinetic Parameter | Reference | Remark |
|---|---|---|---|
| SOD | $k_1 = 2.35 \times 10^9\ M^{-1}s^{-1}$ | [111,113] | Bovine, $T \sim 25\ °C$ |
| $CAT_{H_2O_2}$ | $K_3 = 80.0 \times 10^{-3}\ M$, $k_3 = 0.587 \times 10^6\ s^{-1}$ | [107] | $[H_2O_2] \leq 200\ nM$ |
| $CAT_{ONOO^-}$ | $k_4 = 1.7 \times 10^6\ M^{-1}s^{-1}$ | [100] | Bovine, $T = 25\ °C$ |
| POD | $K_5 = 30 \times 10^{-6}\ M$, $k_5 = 320\ s^{-1}$ | [108] | $t \leq 100\ ms$ |

Since we wish to investigate the kinetics of the species truly originating from superoxide anions, i.e., without an external source of hydrogen peroxide, peroxynitrite, peroxynitrous acid, hypochlorous acid, hydroxyl radicals or nitrogen dioxide, all initial concentrations, except of $[O_2^{\bullet -}]_0$, $[^{\bullet}NO]_0$, $[H^+]_0$, $[Cl^-]_0$, $[CAT]_0$, $[SOD]_0$ and $[POD]_0$ are set to zero. The nonzero initial concentrations are taken from appropriate sources. We assume $pH = 7$, i.e., $[H^+]_0 = 10^{-7}\ M$. The initial concentrations of the ROS and RNS are summarized in Table 3. It should be noted that the initial concentration of nitric oxide ([119]) is an experimental value measured at the surface of a stimulated endothelial cell ($T = 37\ °C$). This is the literature value found for conditions that best resemble ours. The same value is also supported by other studies: for macrophage cells, the (steady-state) concentration of nitric oxide in the vicinity of the cells was ~ $1\ \mu M$ [65,103,104]. The initial concentration of superoxide anions ([64]) is the steady-state concentration of superoxide anions in the extracellular compartment of macrophage cells, which was found to be in the order of $nM$.

**Table 3.** Initial concentrations of the ROS and RNS.

| Species | Initial Concentration (M) | Reference | Remark |
|---|---|---|---|
| $O_2^{\bullet -}$ | $10^{-9}$ | [64] | |
| $H^+$ | $10^{-7}$ | | Assigned |
| $H_2O_2$ | 0 | | Assigned |
| $^{\bullet}NO$ | $10^{-6}$ | [119] | In stimulated endothelial cells |
| $ONOO^-$ | 0 | | Assigned |
| $Cl^-$ | 0.140 | [120] | |
| $HOCl$ | 0 | | Assigned |
| $^{\bullet}OH$ | 0 | | Assigned |
| $ONOOH$ | 0 | | Assigned |
| $^{\bullet}NO_2$ | 0 | | Assigned |

For the initial concentration of the enzymes in the extracellular compartment of cancer cells, no experimental values could be found. How this issue is dealt with will be described in the results section (Section 5).

It should be noted that the parameter values used in this study are taken from experiments performed under conditions that most likely deviate in one or many aspects from the true *in vitro* or *in vivo* conditions in the extracellular compartment of a cancer cell. Having this in mind, we believe these parameter values are still a good starting point for the purpose of these numerical calculations.

### 4.2. Software and details about the calculations

The numerical calculations are performed in MATLAB. Due to significant differences in time scales, we use the solver ode23 s to solve the set of rate equations.

We vary the initial concentration of catalase according to $0 \leq [CAT]_0 \leq [CAT]_{max}$ (the value of $[CAT]_{max}$ will be discussed in Section 5.1), using the concentration interval step size $\Delta[CAT]_0 = 0.01[CAT]_{max}$. Furthermore, we use $t_f = 10\ s$ ($t_f$ denoting the final time) and $\Delta t = 10^{-6}\ s$ ($\Delta t$ denoting the time interval step size) for $0 \leq t \leq 1$, and $\Delta t = 10^{-3}\ s$ for $1 < t \leq 10$ in the calculations.

## 5. Results and discussion

### 5.1. Revealing the dominant pathway for hydroxyl radical generation

Since we cannot find experimental values for the concentration of the different enzymes in the extracellular compartment of cancer cells, we first perform a model analysis to assign feasible values for these parameters.

Information about the fate of superoxide anions, in particular which pathway contributes the most to the generation of hydroxyl radicals under the given conditions, can be extracted by considering the rates by which superoxide anions are consumed in the different pathways. The first reaction in each pathway is given by reactions (1) (the hypochlorous acid pathway) and (2) (the nitric oxide/peroxynitrite pathway). For short times, the rate, $v_1$ and $v_2$, of superoxide anion consumption in each of these reactions is:

$$v_1 = -k_1[O_2^{\bullet -}]_0[SOD]_0,$$
$$v_2 = -k_2[O_2^{\bullet -}]_0[{}^\bullet NO]_0.$$

The ratio of the initial rates is hence

$$\frac{v_2}{v_1} = \frac{k_2[{}^\bullet NO]_0}{k_1[SOD]_0}.$$

Inserting the values of $k_1, k_2$ and $[{}^\bullet NO]_0$ (see Tables 1–3) yields

$$\frac{v_2}{v_1} \sim \frac{10^{-5}}{[SOD]_0}.$$

From this ratio of the initial consumption rates of superoxide anions in both pathways, we can see that the value of $[SOD]_0$ (which is unknown) defines which one of three different regimes of the combined reaction network of the hypochlorous acid- and the nitric oxide/peroxynitrite pathway we are considering. Indeed, since the rate equation for SOD-catalyzed dismutation of superoxide anions is not given by a Michaelis-Menten equation (see [113]), we are not constrained by the condition $[SOD]_0 \ll [O_2^{\bullet -}]$, but we can explore the whole regime of SOD-concentrations. The SOD-defined pathway regimes, according to this model and initial conditions, are:

1. The nitric oxide/peroxynitrite pathway regime occurs for $[SOD]_0 \lesssim 10^{-7}$ M,
$$([SOD]_0 = 10^{-7} M \Rightarrow v_2 \sim 100 v_1).$$
2. The combined pathway regime occurs for $10^{-7} M \lesssim [SOD]_0 \lesssim 10^{-3} M$.
3. The hypochlorous acid pathway regime occurs for $[SOD]_0 \gtrsim 10^{-3}$ M,
$$([SOD]_0 = 10^{-3} M \Rightarrow v_1 \sim 100 v_2).$$

If we assume that the SOD-molecules are solid spheres (in such a case their radius is in the order of $r = 16$ Å- see Appendix A.2), the volume that one SOD-molecule is occupying is given by a cube with the side length $2r$. The physical limit for the concentration of SOD is thus ($N_A$ = Avogadro's constant):

$$[SOD]_{max} = \frac{1}{N_A}\frac{1}{8r^3} \sim 50 \times 10^{-3} \, M.$$

Thus, in order for the hypochlorous acid pathway to be dominant under the given conditions, the concentration of SOD has to be of the same order of magnitude as the physically maximal concentration of SOD. This is of course not feasible since the molecules must be in motion to collide and react. Thus, in this reaction network, the hypochlorous acid pathway will never be the dominant pathway. Moreover, since the generation of hydroxyl radicals in the hypochlorous acid pathway occurs through the reaction (i.e., reaction (6) in the reaction network in Section 3.2)

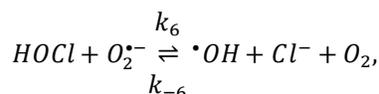

$$HOCl + O_2^{\bullet -} \underset{k_{-6}}{\overset{k_6}{\rightleftharpoons}} {}^\bullet OH + Cl^- + O_2,$$

there will not be a significant generation of hydroxyl radicals for this pathway, even in the regime where it is dominant. This can be seen by making an analysis of the rate of formation of hydroxyl radicals, which is given by

$$v'_6 = -v_6 = k_6[HOCl][O_2^{\bullet -}],$$

where $v_6$ is the rate of superoxide anions consumption in reaction (6). Since the superoxide anion is a precursor of hypochlorous acid, there is a trade-off between a high concentration of hypochlorous acid and a high concentration of superoxide anions. If $[CAT]_0 = 0$, then reaction (3) does not occur. If furthermore the nitric oxide/peroxynitrite pathway is neglected, i.e., reaction (2) is removed, then all superoxide anions will form hydrogen peroxide. Thus, the yield of hypochlorous acid is maximal. Furthermore, $k_{-6} = 0$ (see Table 1). If we assume that hypochlorous acid is formed directly from superoxide anions (i.e., we neglect the intermediate reactions and the kinetics of those), due to the stoichiometry (in reactions (1) and (5)), we know that when hypochlorous acid exists in a certain concentration $[HOCl] = x$, the concentration of superoxide anions is $[O_2^{\bullet-}] = c_0 - 2x$, where $c_0 = [O_2^{\bullet-}]_0$. We are for now not interested in the rate of formation of hypochlorous acid, but merely note that its finite value will decrease the overall rate of the formation of hydroxyl radicals from the hypochlorous acid pathway. Instead, assuming an instant formation of hypochlorous acid from superoxide anions, with resulting concentrations $[HOCl] = x$ and $[O_2^{\bullet-}] = c_0 - 2x$, we now want to investigate the maximal rate of formation of hydroxyl radicals under the given assumptions. With the introduced denotations, the rate of hydroxyl radical formation reads:

$$v'_6(x) = k_6 x(c_0 - 2x) = k_6 x c_0 - 2k_6 x^2.$$

Furthermore,

$$\frac{dv'_6}{dt} = k_6 c_0 - 4k_6 x,$$

and

$$\frac{d^2 v'_6}{dt^2} = -4k_6 < 0.$$

The maximum (since $\frac{d^2 v'_6}{dt^2} < 0$), of the rate of formation of hydroxyl radicals is thus given by

$$\frac{dv'_6}{dt} = 0 \Rightarrow k_6 c_0 - 4k_6 x_{max} = 0 \Leftrightarrow x_{max} = \frac{c_0}{4}.$$

Hence, the function $v'_6$ has a maximum, $v'_{6,max}$, at

$$[HOCl]_{max} = \frac{[O_2^{\bullet-}]_0}{4}.$$

The ratio between the rate of superoxide anions consumption in reaction (1) and reaction (6) is then (the values of $k_1$ and $k_6$ are found in Tables 1 and 2)

$$\frac{v_1}{v_{6,max}} = \frac{4k_1[SOD]_0}{k_6[O_2^{\bullet-}]_0} \sim \frac{10^3[SOD]_0}{[O_2^{\bullet-}]_0}.$$

In order for these two reactions to occur at approximately the same rate (given that $[O_2^{\bullet-}]_0 = 10^{-9}\ M$, see Table 3), $[SOD]_0 \sim 10^{-12}\ M$. In such a case, in order for reactions (1) and (2) to occur at approximately the same rate, $[^{\bullet}NO]_0$ has to be several orders of magnitude smaller than the value reported in Table 3. In fact, $[^{\bullet}NO]_0 \sim 10^{-13}\ M$.

For $[SOD]_0 \sim 10^{-3}\ M$ (the hypochlorous acid pathway regime), $v_1 \sim 10^9 v_{6,max}$, i.e., the generation of hydroxyl radicals from reaction (6) is negligible. Hence, the kinetics of the reaction network suggests that the hypochlorous acid pathway regime is insignificant in terms of the possibility to reactivate hydroxyl radical generation.

For the second regime, where both pathways come into play, the same reasoning applies; there will not be a significant amount of hydroxyl radicals generated from the hypochlorous acid pathway and any superoxide anions going into the hypochlorous acid pathway will reduce the total hydroxyl radical production of the reaction network. Thus, the interesting regime to investigate is the first regime where the nitric oxide/peroxynitrite pathway operates alone.

Regarding the nitric oxide/peroxynitrite pathway alone (i.e., $[SOD]_0 \lesssim 10^{-7}\ M$) the only enzyme-catalyzed reaction (with catalase) is not given as a Michaelis–Menten mechanism (see [107]), so in this case, it should be feasible to use the physically maximal concentration of catalase as $[CAT]_{max}$ (see Section 3.1) to explore the whole regime of possible catalase concentrations. We thus use the concentration of catalase when the whole extracellular compartment is maximally filled with catalase as our default reference value. This concentration is found in the same manner as in the case of SOD. The radius is now $r = 25\ Å$ (see Appendix A.2), which yield $[CAT]_{max} \sim 10^{-2}\ M$. Like in the case of SOD, this does obviously not correspond to a realistic scenario since there will be no space for the molecules to collide, but in lack of an experimental value of the true catalase concentration, we wish to explore the whole regime of physically possible catalase concentrations.

### 5.2. The catalase-dependence of the hydroxyl radical generation in the nitric oxide/peroxynitrite pathway

Since we showed in the previous section that virtually no hydroxyl radicals are generated from the hypochlorous acid pathway in the reaction network considered (even when the condition $[SOD] \gtrsim 10^{-3}\ M$ is fulfilled), we choose to exclude the hypochlorous acid pathway from the numerical calculations, and instead we focus solely on the nitric oxide/peroxynitrite pathway. This pathway consists of reactions (2), (4), (7) and (8). The effect of varying catalase concentrations on the (maximal) production of hydroxyl radicals, can be seen in Figure 4. Note that in regions where the kinetics is highly sensitive to the initial concentration of catalase, the discrete set of initial catalase concentrations may give rise to a discontinuous appearance of the plot representing the corresponding hydroxyl radical concentration.

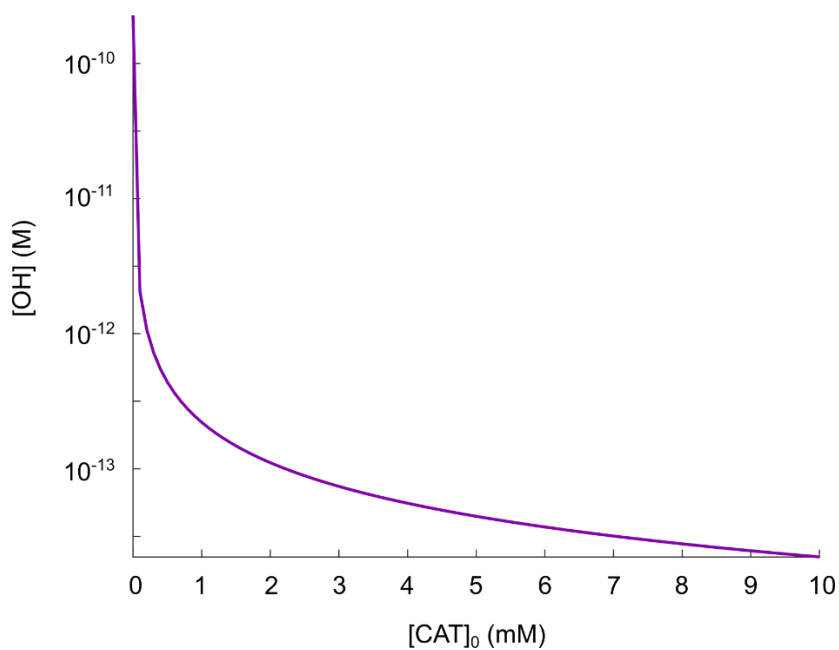

**Figure 4.** Maximal concentration of hydroxyl radicals as a function of catalase concentration in the regime $0 \leq [CAT]_0 \leq 10^{-2}\ M$.

Following the theory in [35,55–61], it must be assumed that $[CAT]_0 = 0\ M$ (see Section 2.2) corresponds to a fully reactivated pathway, i.e., hydroxyl radicals should be produced in such an amount that it is sufficient to induce apoptosis. The result in Figure 4 shows that this occurs at $[^\bullet OH] = 2.3 \times 10^{-10}\ M$. Comparing the concentration of generated hydroxyl radicals at $[CAT]_0 = 0\ M$ with the experimental results in refs. [105,106] (i.e., critical hydroxyl radical concentration for apoptosis-induction, $[^\bullet OH] = 5 \times 10^{-6}\ M$, as mentioned in Section 3.1), we can see that when hydroxyl radicals are added externally, a concentration almost four orders of magnitude higher is required to induce apoptosis. This fact should raise some questions about the possibility to explain CAP-induced apoptosis as a consequence of a reactivation of the nitric oxide/peroxynitrite pathway in the extracellular compartment (since even with no catalase present, there will not be a high enough hydroxyl radical concentration). However, to be able to analyze this theory further, we make the assumption that due to diffusion, as well as the very reactive nature of hydroxyl radicals, adding hydroxyl radicals in the concentration range $10^{-6}\ M$ externally could correspond to a hydroxyl radical concentration in the range $10^{-10}\ M$ close to the cell membrane (where catalase is located). Thus, we assume for now that $[^\bullet OH] = 2.3 \times 10^{-10}\ M$ is sufficient to cause apoptosis-induction.

As can be seen in Figure 4, there seem to be an inverse dependence of $\left|\frac{\Delta[OH^\bullet]}{\Delta[CAT]_0}\right|$ on $[CAT]_0$, i.e., the difference in the amount of generated hydroxyl radicals for two different catalase concentrations, is larger at smaller catalase

concentrations. Thus, we see that in the regime $0 \leq [CAT]_0 \lesssim 10^{-3}\ M$, there is a significant reduction of generated hydroxyl radicals as a result of an increased catalase concentration. For $10^{-3} \lesssim [CAT]_0 \leq 10^{-2}\ M$, on the other hand, the effect of an increased catalase concentration is becoming less significant with respect to an additional reduction of the amount of generated hydroxyl radicals.

Considering the protective effect of catalase, the level of protection (i.e., compared to $[CAT]_0 = 0\ M$) is roughly:

- Two orders of magnitude reduction of $[OH^\bullet]$: $[CAT]_0 = 10^{-4}\ M$.
- Three orders of magnitude reduction of $[OH^\bullet]$: $[CAT]_0 = 10^{-3}\ M$.

Based on these findings, we think that the regime $0 \leq [CAT]_0 \lesssim 10^{-4}\ M$ can be considered to be the most important in the context of catalase-dependent hydroxyl radical generation from the nitric oxide/peroxynitrite pathway and we conclude that it is reasonable to assume that a catalase concentration of $10^{-4}\ M$ can be considered sufficient to protect against hydroxyl radical generation. To analyze this regime further, especially with an increased resolution with respect to $[CAT]_0$, additional calculations for the regime $0 \leq [CAT]_0 \leq 10^{-4}\ M$ are performed (see Figure 5).

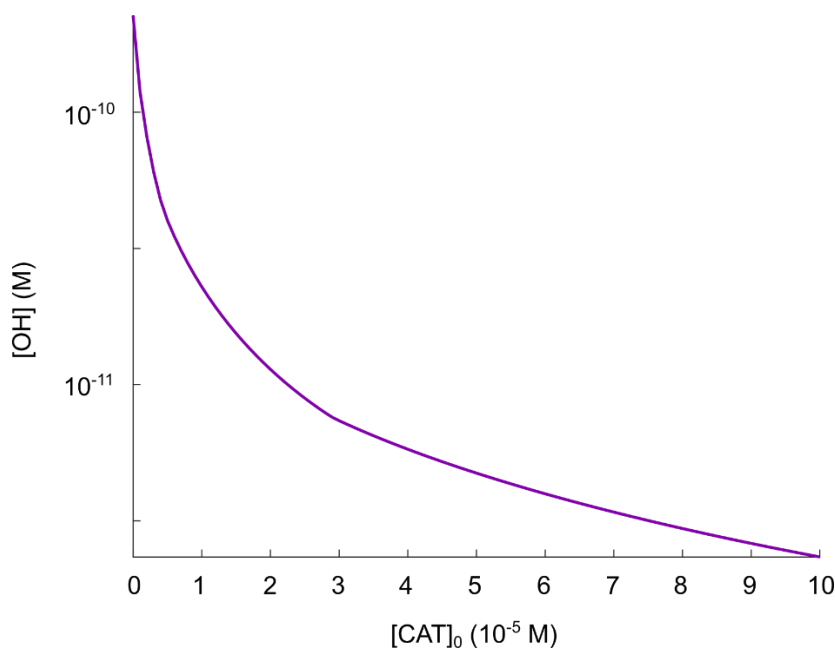

**Figure 5.** Maximal concentration of hydroxyl radicals as a function of catalase concentration in the regime $0 \leq [CAT]_0 \leq 10^{-4}\ M$.

As already observed in Figure 4, $[CAT]_0 = 10^{-4}\ M$ reduces the (maximal) concentration of hydroxyl radicals by about two orders of magnitude, but due to the inverse dependence of $\left|\frac{\Delta[OH^\bullet]}{\Delta[CAT]_0}\right|$ on $[CAT]_0$, a major part of the catalase has to be inactivated to achieve a hydroxyl radical concentration close to that of a fully reactivated pathway. Even for an increase (of the hydroxyl radical concentration) of about one order of magnitude (see Figure 5), about 90% of the initial catalase (i.e., $[CAT]_0 = 10^{-4}\ M$) has to be inactivated (i.e., the effective catalase concentration has to be lowered from $10^{-4}\ M$ to $10^{-5}\ M$). In this sense, the catalase-dependence of the hydroxyl radical generation can be compared to a delta-function, or a switch; the hydroxyl radical generation is either "on" or "off" (almost). This is better seen in Figure 4 (in the regime $0 \leq [CAT]_0 \lesssim 10^{-4}\ M$).

Having analyzed the catalase-dependent kinetics of the hydroxyl radical generation from the nitric oxide/peroxynitrite pathway by numerical calculations, our main observations and conclusions are:

- The maximal concentration of hydroxyl radicals generated from this pathway (i.e., when it is fully reactivated, see Figure 4 at $[CAT]_0 = 0\ M$), is four order of magnitudes lower than the concentration required to induce apoptosis when hydroxyl radicals are added externally ([105,106]). Thus, it seems unlikely that this pathway will account for the anti-cancer effect of CAP.

- If it is assumed that the amount of generated hydroxyl radicals from the fully reactivated nitric oxide/peroxynitrite pathway still is sufficient to cause apoptosis-induction, the level of protection of different catalase concentrations can be analyzed. From the resulting plot (see Figure 4) of the hydroxyl radical concentrations for various catalase concentrations (spanning from zero to the physically maximum amount of catalase), we conclude that $[CAT]_0 = 10^{-4} M$ seem to be a reasonable assumption of the amount of catalase required to protect from hydroxyl radical generation. This is because this catalase concentration reduces the amount of generated hydroxyl radicals by approximately two orders of magnitude compared to a fully reactivated pathways (i.e., when $[CAT]_0 = 0 M$), see Figure 4. For $[CAT]_0 > 10^{-4} M$, on the other hand, the additional protection of an increased catalase concentration is less profound. In Figure 4 it can indeed be seen that in this regime, the slope of the curve is significantly less than for $[CAT]_0 < 10^{-4} M$.
- If it is assumed that $[CAT]_0 = 10^{-4} M$ does protect the cancer cells sufficiently well from apoptosis-induction by hydroxyl radicals, it can be seen in Figures 4 and 5 that a massive decrease of the catalase concentration is required in order to reactivate the hydroxyl radical generation. Indeed, about 90% of the catalase has to be inactivated in order to increase the concentration of hydroxyl radicals with one order of magnitude. This is due to the nonlinear behavior of the kinetics of the nitric oxide/peroxynitrite pathway; the resulting hydroxyl radical generation is very sensitive in the regime $0 \leq [CAT]_0 < 10^{-4} M$ and thus, a small difference in the input (i.e., the catalase concentration) causes a large difference in the output (i.e., the hydroxyl radical concentration).

In order to further evaluate the plausibility of the theory in refs. [37,55–60], where the catalase inactivation is caused by singlet oxygen, we now want to compare these results with an analysis of the extent of catalase inactivation caused by singlet oxygen (see Section 5.3).

### 5.3. Catalase inactivation by primary and secondary singlet oxygen

An important aspect of the theory investigated (see refs. [37,55–60]) is by which means CAP causes catalase inactivation (and subsequently, a reactivation of the hypochlorous acid- and the nitric oxide/peroxynitrite pathway). Here, it is postulated that the inactivation of catalase occurs through interaction with singlet oxygen. In this section, we are investigating the plausibility of this aspect of the theory.

In ref. [102], it was found that catalase is susceptible to oxidative modification and damage (with loss of activity) when exposed to singlet oxygen. The rate of catalase inactivation was dependent on the concentration of singlet oxygen. In ref. [101] it was furthermore shown that the rate constant for the loss of catalase activity induced by singlet oxygen is $k_{10} = 2.5 \times 10^7 M^{-1}s^{-1}$. The loss of enzymatic activity of catalase, $\{CAT\}$, could thus be expressed as

$$\frac{d\{CAT\}}{dt} = -k_{10}[^1O_2]\{CAT\},$$

where $[^1O_2] = 5 pM$ was the used steady-state concentration. Since catalase has four active centers per catalase molecule, the corresponding loss of enzyme activity expressed in terms of $[CAT]$ is

$$\frac{d[CAT]}{dt} = -\frac{k_{10}}{4}[^1O_2][CAT].$$

We assume that the relationship is valid also for $[^1O_2] \neq 5 pM$, and note that the steady-state concentration of catalase is independent of $k_{10}$; only the timescale to reach the steady-state is dependent of the magnitude of $k_{10}$.

*5.3.1. Primary singlet oxygen*

First, we evaluate the effect on the resulting concentration of (active) catalase for various concentrations of "primary" singlet oxygen. It has been found that for various CAP sources, the concentration of singlet oxygen in solution is within the range $10^{-6} \leq [^1O_2]_0 \leq 10^{-5} M$ [121]. In order to definitely cover this concentration range, we are analyzing the regime $0 \leq [^1O_2]_0 \leq 10^{-4}\ M$. We furthermore use $t_f = 10^{-1}\ s, \Delta t = 10^{-7}\ s, [CAT]_0 = 10^{-4}\ M$ (see Section 5.2) and $\Delta[^1O_2]_0 = 10^{-6}\ M$ in the calculation.

The process of catalase inactivation by singlet oxygen is fast, but it is not effective in causing a massive catalase inactivation. Note that it requires four singlet oxygen molecules to completely inactivate one catalase molecule. The resulting (steady-state) catalase concentration as a function of (initial) $[^1O_2]$ is shown in Figure 6. We can see that the inactivation of catalase (in the singlet oxygen regime of interest), is not profound. Indeed, even for $[^1O_2]_0 = 10^{-4}\ M$, the resulting catalase concentration is 75% of its initial value. Comparing with Figure 5, we see that the reactivation of the hydroxyl radical generation in this regime is negligible; the increase in the resulting hydroxyl radical concentration is much less than one order of magnitude when $[CAT]_0 = 7.5 \times 10^{-4}\ M$ instead of $[CAT]_0 = 10^{-4}\ M$.

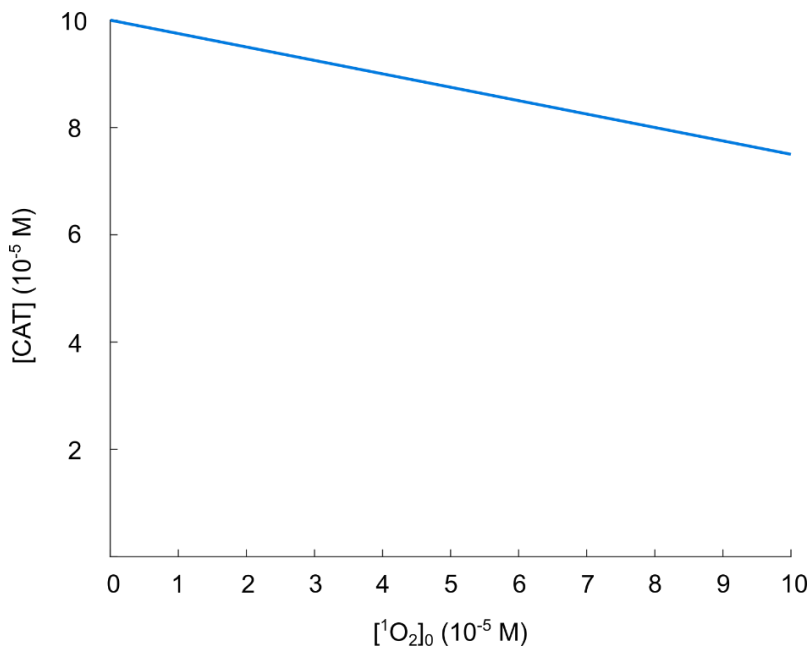

**Figure 6.** Steady-state concentration of catalase as a function of singlet oxygen concentration.

*5.3.2. Secondary singlet oxygen*

It has been suggested that "secondary" singlet oxygen is generated in a sequence of reactions starting with the formation of peroxynitrite from hydrogen peroxide and nitrite [57,59,122]. Subsequently, hydrogen peroxide and peroxynitrite may interact and generate singlet oxygen [34,42,123,124]. This reaction is experimentally verified, but the exact mechanism is yet not fully known. In ref. [125], the results identify the generation of singlet oxygen from peroxynitrate (which is generated from peroxynitrite) through the decomposition reaction.

$$O_2NOO^- \xrightarrow{k} NO_2^- + {}^1O_2.$$

Thus, the formation of singlet oxygen is not (directly) dependent on the concentration of hydrogen peroxide. The total generation of singlet oxygen from peroxynitrite can be written:

$$ONOO^- \xrightarrow{k_{11}} {}^1O_2,$$

where the rate constant was found to be $k_{11} \sim 6.7 \times 10^{-3}\ s^{-1}$ at $pH = 7.2$ and room temperature. Thus, the rate of singlet oxygen generation from decomposition of peroxynitrite is given by

$$\frac{d[^1O_2]}{dt} = k_{11}[ONOO^-].$$

Here, we evaluate the effect on the resulting concentration of (active) catalase for various concentrations of peroxynitrite (and thus, different concentrations of generated "secondary" singlet oxygen) in the regime $0 \leq [ONOO^-]_0 \leq 1\ M$. We use $t_f = 10^0\ s$, $\Delta t = 10^{-6}\ s$, $[CAT]_0 = 10^{-4}\ M$ (see Section 5.2) and $\Delta[ONOO^-]_0 = 10^{-2}\ M$ in the calculation. Furthermore, we assume a certain initial reduction of the catalase concentration to account for the effect of primary singlet oxygen in CAP (as the theory in refs. [37,55–60] describes), thus; $[CAT]_0 = p \times 10^{-4}\ M$, where $p = 0.75$ is the value of the maximal catalase inactivation effect of primary singlet oxygen found in Section 5.3.1.

The resulting (steady-state) catalase concentration as a function of (initial) $[ONOO^-]$ is presented in Figure 7. As can be seen, even at $[ONOO^-]_0 = 1\ M$, which is a concentration of nine orders of magnitude higher than expected in the extracellular compartment of cancer cells, the resulting catalase concentration is $6.1 \times 10^{-4}\ M$, i.e., about 61% of its initial value (the effect of primary singlet oxygen included). Comparing with Figure 5, we see that this catalase concentration does not correspond to a significant reactivation of the generation of hydroxyl radicals.

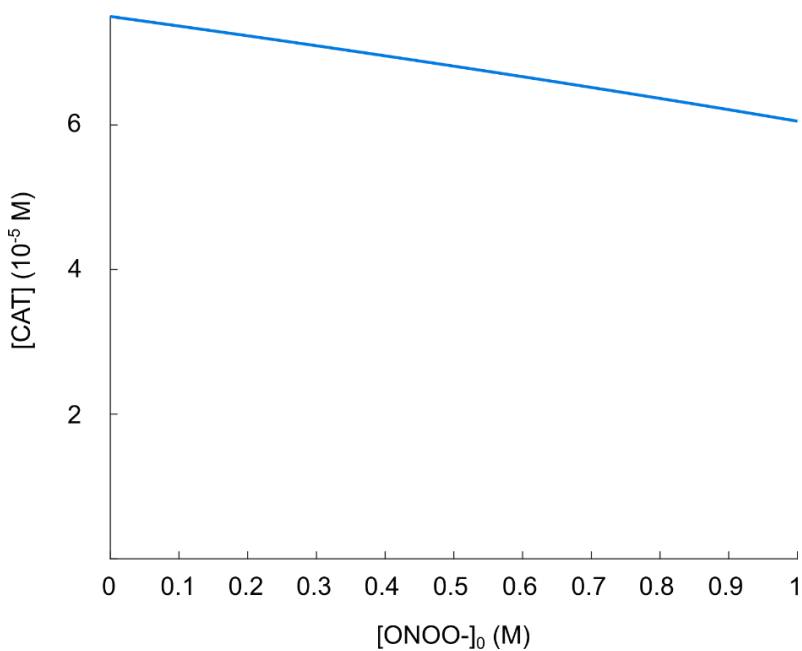

**Figure 7.** Steady-state concentration of catalase as a function of peroxynitrite concentration.

*5.3.3. Summary of the effect of singlet oxygen in the reactivation of hydroxyl radicals in the nitric oxide/peroxynitrite pathway*

In summary; for a catalase concentration that is high enough to cause a significant reduction of the concentration of hydroxyl radicals generated from the nitric oxide/peroxynitrite pathway - and hence, protect against hydroxyl radical generation - singlet oxygen (from CAP and, subsequently, generated in the solution) will most certainly not inactivate catalase to the extent necessary for a reactivation of the nitric oxide/peroxynitrite pathway. This conclusion provides another indication that this theory seems unlikely to account for the underlying cause of the anti-cancer effect of CAP.

It should furthermore be mentioned that there are contradictory results concerning the effect of singlet oxygen on catalase activity as well; there are also cases where catalase from different sources is oxidized by singlet oxygen to active, more acidic enzyme conformers [126]. Here, the catalytic rate constant was almost identical for the non-oxidized and oxidized form of catalase under physiological conditions. Thus, the oxidized catalase would be as efficient as the non-oxidized and the effect of singlet oxygen, in the context of enzyme inactivation, will be none.

## 5.4. The impact of carbon dioxide-catalyzed decay of peroxynitrite on the hydroxyl radical generation in the nitric oxide/peroxynitrite pathway

As a final contribution to the evaluation of the catalase-dependent reactivation of hydroxyl radical generation in the nitric oxide/peroxynitrite pathway, we extend the previous reaction network of a closed system to include a major open system effect regarding the generation of hydroxyl radicals from peroxynitrite.

In the presence of a high concentration of carbon dioxide, which is the case in a biological system, the main route of peroxynitrite decay is through the reaction [127,128]

$$ONOO^- + CO_2 \xrightarrow{k_9} ONOOCOO^-,$$

where $k_9 = (5.8 \pm 0.2) \times 10^4\ M^{-1}s^{-1}$ at $T = 37°C$ [127]. Thus, the rate of peroxynitrite decay is given by

$$\frac{d[ONOO^-]}{dt} = -k_9[ONOO^-][CO_2].$$

Here, we analyze how the catalase-dependent generation of hydroxyl radicals in the nitric oxide/peroxynitrite pathway is affected by a physiological concentration of carbon dioxide in the regime $0 \leq [CAT]_0 \leq 10^{-4}\ M$ (see Section 5.2), and compare with the result presented in Figure 5. As previously, we consider the maximal concentration of hydroxyl radicals.

We assume that the concentration of carbon dioxide is constant (since $[CO_2] \gg [ONOO^-]$) and equal to $[CO_2] = 10^{-3}\ M$, which is a concentration of the correct order of magnitude to represent a physiological level of carbon dioxide, see Appendix A.3. Furthermore, we note that the first reaction in the nitric oxide/peroxynitrite pathway, reaction (2), occurs very fast and results in a concentration of peroxynitrite in the (sub)$nM$-regime ($[ONOO^-]_{max} = 10^{-9}\ M$ since $[O_2^{\bullet-}]_0 = 10^{-9}\ M$). To simplify the calculations, we assume that the maximal amount of peroxynitrite is already formed before any other reaction takes place, i.e., $[ONOO^-]_0 = 10^{-9}\ M$. We use $t_f = 1\ s$ and $\Delta t = 10^{-6}\ s$ in the calculation.

The result of the calculation is presented in Figure 8. As can be seen, in the presence of a physiological concentration of carbon dioxide, the generation of hydroxyl radicals is very low at all catalase concentrations in the regime $0 \leq [CAT]_0 \leq 10^{-4}\ M$. Indeed, even at $[CAT]_0 = 0\ M$ we can see that $[^\bullet OH] = 6.6 \times 10^{-12}\ M$. This value should be compared with $[^\bullet OH] = 2.3 \times 10^{-12}\ M$ at $[CAT]_0 = 10^{-4}\ M$ in Figure 5. Thus, in a realistic biological system, it is even less likely that cells need to protect themselves from hydroxyl radicals by extracellular catalase (cf. Section 5.2). This result speaks against the possibility that this pathway should be important to explain the underlying cause of the anti-cancer effect of CAP.

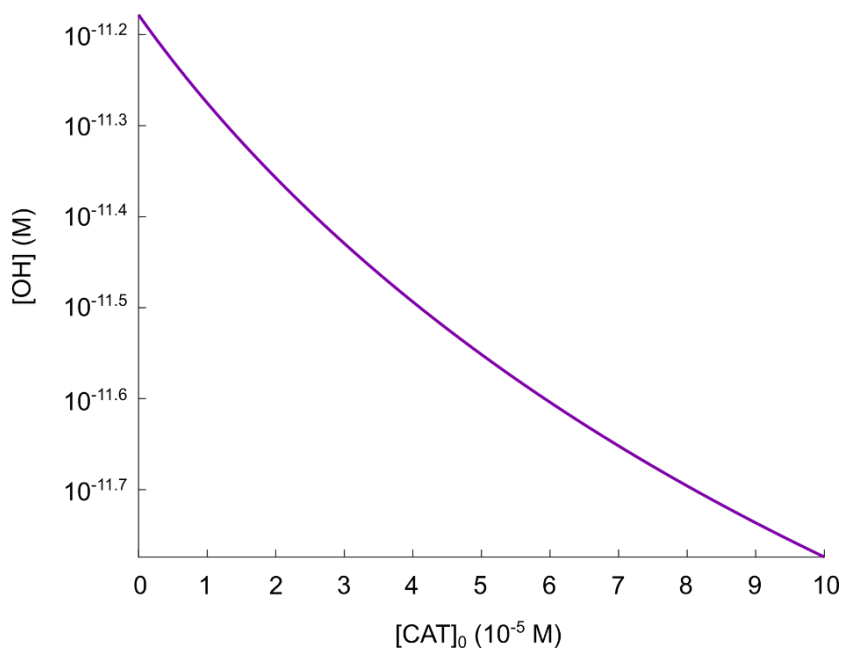

**Figure 8.** Maximal concentration of hydroxyl radicals as a function of catalase concentration in the regime $0 \leq [CAT]_0 \leq 10^{-4}\,M$.

### 5.5. Limitations of the model and potential implications

In our mathematical model of the catalase-dependent kinetics of the reaction network of the hypochlorous acid- and the nitric oxide/peroxynitrite pathway, we mainly consider a closed system as it would appear in a well-mixed reaction vessel. More about the justification of this restriction can be found in Section 3.1. The analysis of the kinetics of the reaction network itself provides information about the *maximal* hydroxyl radical generation from the hypochlorous acid- and nitric oxide/peroxynitrite pathways under the given conditions, and how the interaction with catalase at various concentrations affects the resulting concentration of hydroxyl radicals. It also reveals what concentrations of the involved ROS, RNS, and enzymes are required for a significant generation of hydroxyl radicals from the hypochlorous acid- and nitric oxide/peroxynitrite pathways. Even though the results strongly indicate that these pathways cannot explain the (selective) anti-cancer effect of CAP, it should be mentioned that in a more realistic, open system of the extracellular compartment, there are a number of effects that potentially could affect the results. These are in particular:

- The spatial dynamics of the involved species.
- The constant generation of some species (superoxide anions and nitric oxide in particular).
- The effect of (more) interfering pathways (which in general will cause a reduction of the generation of hydroxyl radicals, because the substrates of hydroxyl radical generation are used in the formation of other products).
- The effect of a possible pH gradient between the extracellular- and intracellular compartments.

First, the fact that a purely kinetic model does not result in a generation of hydroxyl radicals from the hypochlorous acid pathway, while experiments suggest that this should be the case, *could* indicate that both *in vitro* and *in vivo*, the spatial dynamics might play an important role. If a fraction of superoxide anions diffuses away from its site of production before reacting with nitric oxide or SOD, it could serve as a reservoir of superoxide anions for hypochlorous acid to react with. However, the generation of hydroxyl radicals from the hypochlorous acid pathway might then not occur in direct vicinity to the cell membrane, which reduces the chances for hydroxyl radicals to cause lipid peroxidation (and hence, apoptosis-induction). Furthermore, other studies in similar systems predict superoxide anions to be consumed within a very narrow spatial regime from its site of production [64,65]. Diffusion of nitric oxide from the site of production would lead to a lower effective concentration close to the cell membrane, thus making it more likely for superoxide anions to enter the hypochlorous acid pathway. Still, the kinetics of this reaction network predicts that [$^{\bullet}NO$] has to decrease by many orders of magnitude and at the same time $[SOD]_0$ must be very low, in order for (a very small amount of) hydroxyl radicals to be generated from the hypochlorous acid pathway. Furthermore, the total concentration of generated hydroxyl radicals from the reaction network will be reduced in such a case.

Second, it can be discussed how the results would be expected to differ if a constant generation of superoxide anions and nitric oxide were included in the model. Considering the time scale (about 200 times longer than the time scale relevant here) to reach a steady-state concentration of nitric oxide in a similar system [104], we can assume that adding such terms to the system of rate equations would not significantly change the results. If so, a constant generation of nitric oxide in the reaction network would obviously increase the fraction of superoxide anions entering the nitric oxide/peroxynitrite pathway. A constant generation of superoxide anions, on the other hand, should not change the qualitative result since also the newly generated superoxide anions would directly enter one of the two initial reactions in the reaction network, thus contributing to the formation of peroxynitrite and hydrogen peroxide, not to the formation of hydroxyl radicals through reaction between hydrogen peroxide and superoxide anions. Thus, we believe that the possible generation of hydroxyl radicals in such a case would still be negligible.

Third, including any of the interfering reactions would - with one exception - only lower the total generation of hydroxyl radicals from the reaction network. The exception is that FASR (which is an apoptosis-inducing membrane-receptor) can be triggered by singlet oxygen in the absence of its genuine ligand [129], a process that activates caspase-8 and enhances the NOX1 activity as well as induces NOS expression [130,131]. Thus, there will be a higher local generation of superoxide anions and nitric oxide. Referring back to the previous discussion, if this will significantly change the result, it will increase the peroxynitrite concentration, thus making the nitric oxide/peroxynitrite pathway stronger. Including reactions where nitric oxide is consumed is also a possible way to decrease the effective concentration of nitric oxide, which would increase the amount of superoxide anions entering the hypochlorous acid pathway. Still, as already mentioned, this causes a lower total concentration of hydroxyl radicals generated from the total reaction network.

Fourth, the effect of a pH gradient between the extracellular and intracellular compartments could change the condition for the nitric oxide/peroxynitrite pathway in such a manner that a higher amount of hydroxyl radicals are generated. However, the pH in the extracellular compartment must be substantially lower than normal to create a significant difference in the result. As an example, for $pH = 6$ and $[CAT]_0 = 0\ M$, the resulting concentration of hydroxyl radicals is $[^{\bullet}OH] = 3.2 \times 10^{-10}\ M$ instead of $[^{\bullet}OH] = 2.3 \times 10^{-10}\ M$ (at $pH = 7$ and $[CAT]_0 = 0\ M$). Nevertheless, the pH-dependence of the hydroxyl radical generation could be interesting to include in future models.

In addition, we should also mention the effect of temperature and viscosity on the rate constants, especially since two important rate constants in the hypochlorous acid pathway ($k_1$ and $k_6$, see Section 4.1) are from studies performed at room temperature (rather than at the relevant temperature of 37 °C). However, it seems unlikely that these rate constants would be increased by many orders of magnitude due to the relatively small increase in temperature. Thus, we believe that the qualitative behavior of the reaction network would not change substantially by adjusting these rate constants to 37 °C. The viscosity in the extracellular compartment is probably higher than the solutions for which the included rate constant values are determined. Since this means a slower diffusion for the reacting species, it leads to less frequent collisions and thus lower rate constants. If this effect is more noticeable for some reactions than others, it could affect the model results.

As mentioned in the introduction (Section 1), there are no quantitative experimental results to compare our theoretical results to. The lack of experimental data on the resulting concentration of hydroxyl radicals for different initial conditions of catalase concentrations, which, in turn, could be related to the apoptosis-induction capacity, prevents us from a direct comparison of the results that our mathematical model predicts, and experimental results from the reaction network at identical conditions. However, regarding our conclusion that these apoptotic pathways are unlikely to contribute to the anti-cancer effect of CAP, it should be mentioned that recently a study on the effect of CAP on the superoxide generating NOX1 in cancer cells has been conducted [132]. Here, it was found that CAP is reducing the amount of active NOX1, which thus leads to a decreased generation of superoxide anions. This result is hence also in contradiction with the theory in refs. [37,55–60].

Regarding the nitric oxide/peroxynitrite pathway, we are not including the potential oxidation of nitric oxide by (active) catalase in this study. It has been excluded as the main effect of catalase on the nitric oxide/peroxynitrite pathway [34], but may be included in future models to rule out that it would have a significant effect on the concentration of nitric oxide. Another aspect that would be more important to include in studying this pathway, is that experiments show that the formation of hydroxyl radicals from peroxynitrous acid should rather be written as

$$ONOOH \rightarrow f(^\bullet OH + {}^\bullet NO_2) + (1-f)(H^+ + NO_3^-),$$

where $0 \leq f \leq 1$ is the fraction of peroxynitrous acid that undergoes radical escape. The yield of hydroxyl radicals from decomposition of peroxynitrous acid, according to the literature, varies within the range $0 - 40\%$ [133–139], i.e., $f \leq 0.4$ whereas in this study, $f = 1$, is assumed. Thus, our results overestimate the generation of hydroxyl radicals also with regard to this aspect.

Regarding the hypochlorous acid pathway and the discrepancy between our results (which predict that hydroxyl radicals are not generated from this pathway) and the experimental findings (where the important role of this pathway is claimed to be established), an aspect that should be brought up in the discussion about this pathway, is that it has been shown that superoxide anions can be generated by two common components of cell culture media: HEPES buffer and (in the presence of light) riboflavin [140]. Thus, for *in vitro* studies in such cell culture media, some of the superoxide anions production might be attributed to the generation of the media and not by the cells themselves. In e.g., refs. [94,96], where the hypochlorous acid pathway was studied, Eagle's Minimal Essential Medium was used for the cell cultures. Since Eagle's Minimal Essential Medium contains riboflavin and since there is nothing mentioned about the light conditions for these experiments, it cannot be ruled out that the results from these studies are affected by the production of superoxide anions by the culture medium. Another aspect regarding this pathway is that possible SOD-inactivation of CAP indicates that the hypochlorous acid pathway might be less important than the nitric oxide/peroxynitrite pathway in the context of apoptosis-induction by CAP. Indeed, since SOD shares the common feature of histidine at the active site with catalase, it is reasonable to assume that also SOD is inactivated by CAP in the same manner as catalase. It has been shown that the histidine residue of SOD is inactivated by singlet oxygen even more effectively than the histidine residue of catalase [101,102]. In such a case, the probability of SOD inactivation by CAP should be even higher than catalase inactivation. The immediate effect of SOD inactivation is that the rate of hydrogen peroxide formation is decreased, whereas the peroxynitrite concentration is increased. Furthermore, the increased concentration of superoxide anions may cause catalase inhibition. In ref. [122] it is verified that inhibition of SOD increases the concentration of superoxide anions, which inhibits catalase and removes the enzymatic dismutation of superoxide anions to hydrogen peroxide, thus inhibiting the hypochlorous acid pathway. It has indeed been shown that inactivation of SOD indirectly inhibits catalase through superoxide anions dependent inhibition [141]. In this study, inactivation of SOD was also found to cause a drastic decrease of hydrogen peroxide.

In ref. [55], another potential effect contributing to hydroxyl radical-induced apoptosis, is mentioned. In the case of aquaporins in the vicinity of inactivated catalase, hydrogen peroxide could be transferred from the extracellular compartment to the intracellular compartment. Here, hydrogen peroxide is claimed to cause depletion of glutathione. It is furthermore claimed that this process makes the cell more sensitive to extracellular hydroxyl radicals. The resulting effect might be that the more catalase that is inactivated, the less amount of hydroxyl radicals is required to induce apoptosis. In such a case, catalase does not have to be inactivated to such a high extent that our results suggest, i.e., our results might have to be adjusted. However, considering how very little effect singlet oxygen has on catalase activity in the relevant regime, we believe that this cannot change our conclusions significantly.

Another critical aspect of the theory of the catalase-dependent apoptotic pathways that is worth mentioning is whether it is fundamentally deterministic or stochastic; it is mentioned in e.g., ref. [60] that only "a few" catalase molecules need to be initially inactivated in order to allow for the generation of secondary singlet oxygen at the sites of inactivated catalase. A dynamical system with only "a few" molecules - at least initially - should be described by an underlying stochastic theory, i.e., a theory where a certain future event only can be determined to occur with a certain probability. However, the experimental results in e.g., refs. [57–60] indicate an underlying deterministic theory since the results are reproducible as long as systems are prepared in the same manner.

Lastly, there are some additional effects, not included in the model, that also could change the results. These are:

- The difference in the catalytic action of membrane-bound enzymes as compared to enzymes that are free in solution.
- The effect of non-equilibrium on the rate constants.
- The potential enzyme inhibition by the products (or other species).
- The pH-dependence of enzyme activity.

These effects are, however, all very difficult to take into account, but may play an important role in *in vivo* systems.

In summary, our model does not capture the full, very complex set of reactions and mechanisms that describes the apoptotic pathways (in refs. [37,55–60]) in a cancer cell *in vitro* or *in vivo*, but the negative results regarding the theory that we aim to reproduce should still be considered as a strong indication that the underlying cause of CAP-induced cancer cell death cannot be understood in terms of catalase-dependent reactivation of these apoptotic pathways. In particular, this is true since we use optimizing conditions for the hydroxyl radical generation and explore very large concentration regimes of those species that act as independent variables in the calculations. We hope that our results will motivate other researchers to further examine this theory, before it can be accepted as a plausible theory of the underlying cause of the anti-cancer effect of CAP.

## 6. Conclusions

In this work, we develop and use a mathematical model to analyze and gain insights in a specific theory, which has been outlined in great detail and is proposed to account for the underlying cause of the (selective) anti-cancer effect of CAP (see refs. [37,55–60]). Explicitly, the model describes the catalase-dependent reaction kinetics of two apoptosis-inducing signaling pathways - the hypochlorous acid pathway, and the nitric oxide/peroxynitrite pathway - occurring in the extracellular compartment of cancer cells. By implementing our model, we investigate the possibility to reactivate the generation of hydroxyl radicals (responsible for apoptosis-induction) from the reaction network of both pathways by inactivation of catalase (which is proposed to be caused by "primary" singlet oxygen contained in CAP, as well as "secondary" singlet oxygen that is subsequently generated). Our analysis suggests that this theory should be critically evaluated in order to be accepted and established as a theory accounting for the underlying cause of the anti-cancer effect of CAP. Here, we list the main findings supporting this claim:

1. The hypochlorous acid pathway is highly unlikely to generate hydroxyl radicals at all at the conditions outlined by the theory. This is due to a very unfavorable kinetics of this pathway with respect to hydroxyl radical formation, causing the rate of generation of hydroxyl radicals to be negligible.
2. The nitric oxide/peroxynitrite pathway may only generate - under conditions that absolutely maximize the yield of hydroxyl radicals - hydroxyl radicals in a concentration that is almost four orders of magnitudes lower than what has been found to be the critical extracellular hydroxyl radical concentration for apoptosis-induction [105,106]. It thus seems unlikely that this pathway will result in apoptosis-induction, whether there is catalase in the extracellular compartment or not.
3. If it is assumed that the amount of generated hydroxyl radicals from the fully reactivated nitric oxide/peroxynitrite pathway is sufficient to cause apoptosis-induction, then a catalase concentration of about 100 µM is required to protect the cells. However, when a physiological concentration of carbon dioxide is introduced into the reaction network of the nitric oxide/peroxynitrite pathway, the effect of catalase in the system is negligible since the generation of hydroxyl radicals is vanishing at all catalase concentrations. Thus, there is no reactivation of the apoptotic pathway at any level of catalase inactivation.
4. If the nitric oxide/peroxynitrite pathway still is assumed to represent a possible apoptotic pathway, then one crucial result still remain; sufficient catalase inactivation (i.e., sufficient with respect to the possibility of pathway reactivation) will most likely not occur by the primary and secondary singlet oxygen possibly contained and generated in the solution after CAP-exposure.

In order to reveal the proper underlying mechanisms, which are likely to be many and different from cell line to cell line, which can sufficiently explain and describe the anti-cancer effect of CAP, we believe that new theories should be proposed, and existing ones should be thoroughly analyzed and evaluated.

# Appendix A.

We present here more detailed information about the rate equations and rate constants used, as well as how the effective radius of enzyme molecules is found.

**Appendix A.1. Rate equations and rate constants**

*Appendix A.1.1. General information*

Reactions (3) and (5) follow Michaelis–Menten kinetics, i.e.,

$$\frac{d[P]}{dt} = k_E [E]_0 \frac{[S]}{K_E + [S]},$$

$$\frac{d[S]}{dt} = -\frac{d[P]}{dt},$$

$$\frac{d[E]}{dt} = 0,$$

where $[S]$ denotes the substrate concentration, $[P]$ denotes the product concentration, $[E]_0$ denotes the initial enzyme concentration, $k_E$ is the catalytic constant for enzyme $E$ and $K_E$ is the Michaelis–Menten constant for enzyme $E$. For a reaction of the type

$$nS + E \xrightarrow{k_E} E + mP,$$

we assume the rate equation

$$-\frac{1}{n}\frac{d[S]}{dt} = \frac{1}{m}\frac{d[P]}{dt},$$

to be valid.

*Appendix A.1.2. Reaction* (1)

In ref. [142] it was found that the activity of (bovine) SOD is relatively independent of $pH$ between 5.3 and 9.5. In ref. [113] it was found that bovine SOD is catalyzing the decay of superoxide anions with a second order rate constant, $k_1 = 2.3 \times 10^9 \ M^{-1}s^{-1}$, at $T = 20 - 25$ °C and $pH = 7.0$, where the rate of superoxide anions decay is given as

$$\frac{d[O_2^{\bullet-}]}{dt} = -k_1[O_2^{\bullet-}][SOD].$$

The rate of formation of hydrogen peroxide should hence follow

$$\frac{d[H_2O_2]}{dt} = \frac{1}{2}k_1[O_2^{\bullet-}][SOD].$$

In ref. [111], it was found that for (bovine) SOD at $T \sim 25$°C and $pH$ between 9.0 and 9.9, the second-order rate constant for superoxide anions decay is $k_1 = 2.4 \times 10^9 \ M^{-1}s^{-1}$. The average of both values for $k_1$ is used in our calculations, i.e., $k_1 = 2.35 \times 10^9 \ M^{-1}s^{-1}$.

*Appendix A.1.3. Reaction* (2)

In ref. [143] the second order rate constant for the decay of superoxide anions at $T = 20 - 25$ °C and $pH = 7.5$ was determined to be $k_2 = (6.7 \pm 0.9) \times 10^9 M^{-1}s^{-1}$. The rate of the decay of superoxide anions hence is

$$\frac{d[O_2^{\bullet-}]}{dt} = \frac{d[^{\bullet}NO]}{dt} = -k_2[O_2^{\bullet-}][^{\bullet}NO],$$

$$\Rightarrow \frac{d[ONOO^-]}{dt} = k_2[O_2^{\cdot-}][{}^{\cdot}NO].$$

In ref. [144] the rate constant for superoxide anions decay was shown to be independent of $pH$ in the range $6.1 - 10.0$ and the rate constant was determined to be $k_2 = (4.3 \pm 0.5) \times 10^9 \, M^{-1}s^{-1}$ (no information on $T$). In ref. [116], the rate constant was found to be $k_2 = 4.8 \times 10^9 \, M^{-1}s^{-1}$ at $T = 25$ °C and in [145] the weighted average of three different experiments gave a value of $k_2 = (1.6 \pm 0.3) \times 10^{10} \, M^{-1}s^{-1}$ (no information on $T$). In ref. [109], a value of $k_2 = 3.2 \times 10^{10} \, M^{-1}s^{-1}$ at $T = 25$ °C was reported. In ref. [109] it was furthermore concluded that the expected rate constant at $T = 37$ °C is $k_2 = 1.7 \times 10^{10} \, M^{-1}s^{-1}$. We use the latter value in our calculations since it is the only value found for the right temperature condition ($T = 37$ °C).

The value of the first order rate constant for the reverse reaction is $k_{-2} = 0.017 \, s^{-1}$ in aqueous solutions at $T = 20$ °C [117] as well as at $T = 25$ °C [116], i.e.,

$$\frac{d[ONOO^-]}{dt} = -k_{-2}[ONOO^-]$$

$$\Rightarrow \frac{d[O_2^{\cdot-}]}{dt} = \frac{d[{}^{\cdot}NO]}{dt} = k_{-2}[ONOO^-].$$

*Appendix A.1.4. Reaction (3)*

The reaction follows Michaelis–Menten kinetics (at $[H_2O_2] \leq 200 \, mM$). Apparent values of $K_3$ and $k_3$ at $pH = 7.0$ and $T = 37$°C are $K_3 = 80.0 \, mM$ and $k_3 = 587000.0 \, s^{-1}$ [107]. Note that in this reaction, it is assumed that the product is oxygen.

*Appendix A.1.5. Reaction (4)*

The reaction was found to be catalyzed by bovine liver catalase with a (second order) rate constant, $k_4 = 1.7 \times 10^6 \, M^{-1}s^{-1}$ at $pH = 7.1$ and $T = 25$ °C [100]. The rate of the decay of peroxynitrite can hence be expressed as

$$\frac{d[ONOO^-]}{dt} = -k_4[ONOO^-][CAT].$$

At $T = 37$ °C, the rate constant of peroxynitrite decay was found to be $2.5 - 3$ times higher than that measured at $T = 25$ °C. However, the authors noted that this result can only serve as an estimation due to some experimental errors. Since the difference in value anyway does not differ by many orders of magnitude, we choose to use the experimental value for $T = 25$ °C in our calculations.

*Appendix A.1.6. Reaction (5)*

The reaction follows Michaelis–Menten kinetics (with $H_2O_2$ as the substrate). The effect of $[Cl^-]$ and $pH$ ($4.4 \leq pH \leq 6.2$) on the $K_5$-value at $T = 37$ °C for hydrogen peroxide of MPO has been studied [146] and it shows that the $K_5$-values decreases with increasing $[Cl^-]$ and $pH$ in a dose-dependent manner. In other words, as $pH$ increases, $K_5$ becomes less sensitive to the concentration of chlorine anions. Hence, for $pH \sim 7$, it should be a valid assumption that $K_5$ changes only very little with changing $[Cl^-]$. For MPO from leukocytes at $pH = 7.2$, $T = 20$ °C, $[H_2O_2] = 100 \, \mu M$ and $[Cl^-] = 200 \, mM$, $k_5 = 320 \, s^{-1}$ per monomer (MPO is a homodimer, i.e., it consists of two monomers, so the effective concentration of MPO is twice the given one) and $K_5 = 30 \, \mu M$ during the first $100 \, ms$ [108] with respect to $H_2O_2$. Thereafter, $k_5$ decreases. For lower concentrations of hydrogen peroxide ($[H_2O_2] < 50 \, \mu M$), $k_5$ is less dependent on time. Thus, for the current calculations ($[H_2O_2] \ll 50 \, \mu M$) it can be justified to assume that $K_5 = 30 \, \mu M$ is a constant in time.

*Appendix A.1.7. Reaction (6)*

At $pH = 5.5$ and $T \sim 20$ °C, the decay of $O_2^{\cdot-}$ in the presence of hypochlorous acid has been determined to occur with a second order rate constant $k_6 = 7.5 \times 10^6 \, M^{-1}s^{-1}$ [110], i.e.,

$$\frac{d[O_2^{\bullet -}]}{dt} = -k_6[O_2^{\bullet -}][HOCl],$$

$$\Rightarrow \frac{d[{}^{\bullet}OH]}{dt} = k_6[O_2^{\bullet -}][HOCl].$$

Since no reported value of $k_{-6}$ could be found and hydroxyl radicals is known to be a very reactive species, thus reacting with almost any other species in its surrounding, we assume that the reverse reaction is of minor importance and $k_{-6}$ is set to 0. (A value $k_{-6} > 0$ will only reduce $[{}^{\bullet}OH]$).

*Appendix A.1.8. Reaction* (7)

The ratio of $k_7$ and $k_{-7}$ is given by

$$K_a = \frac{[ONOO^-][H^+]}{[ONOOH]} = \frac{k_{-7}}{k_7} = 10^{-pK_a} \, M^{-1}.$$

The $pK_a$-value at $T = 25\,°C$ is $6.5 - 6.8$ [118,147]. To optimize the production of hydroxyl radicals, we chose $pK_a = 6.8$ in our calculations.

Diffusion-controlled second-order rate constants of protonation of a wide range of compounds are within $(0.35 - 18) \times 10^{10} \, M^{-1}s^{-1}$ [112]. We choose $k_7 = 10^{10} \, M^{-1}s^{-1}$ in our calculations.

*Appendix A.1.9. Reaction* (8)

The rate constants in reaction (8) has been determined to $k_8 = 10^{-4} - 0.6 \, s^{-1}$ at $T = 25\,°C$ [114] and $k_{-8} = (4.5 \pm 1.0) \times 10^9 \, M^{-1}s^{-1}$ ($pH = 9.5$ and no information about $T$) [115], i.e.,

$$\frac{d[ONOOH]}{dt} = k_{-8}[{}^{\bullet}NO_2][{}^{\bullet}OH] - k_8[ONOOH],$$

and

$$\frac{d[{}^{\bullet}OH]}{dt} = -k_{-8}[{}^{\bullet}NO_2][{}^{\bullet}OH] + k_8[ONOOH].$$

To find the upper limit of the hydroxyl radicals production from the nitric oxide/peroxynitrite pathway, we use the largest possible value for $k_8$, i.e., $k_8 = 0.6 \, s^{-1}$, and the smallest possible value for $k_{-8}$, i.e., $k_{-8} = 3.5 \times 10^9 \, M^{-1}s^{-1}$.

**Appendix A.2. Enzyme effective radius**

The effective radius of the enzymes are found by the following procedure: The pdb-file for each respective enzyme is downloaded on the protein data bank www.rcsb.org. This file is then uploaded into the protein volume calculating program "Voss Volume Voxelator" (http://3vee.molmovdb.org/index.php), which calculates the effective radius of the enzyme.

*Appendix A.2.1. Catalase*

The crystal structure of catalase (homo sapiens) was taken from [148]. The calculations yield the effective radius equal to 25.10 Å. This yields a maximum concentration of 13.1 $mM$.

*Appendix A.2.2. SOD*

The crystal structure of SOD (homo sapiens) was taken from [149]. The calculations yield an effective radius of 15.70 Å. This yields a maximum concentration of 53.6 $mM$.

**Appendix A.3. Concentration of carbon dioxide**

The partial pressure of carbon dioxide in human alveolar has been found to be $p_{CO_2} = (4.0 - 9.3) \times 10^3\ Pa$ [150], which according to Henry's law yields $[CO_2] = Hp$, where $H = 3.4 \times 10^{-2} M\ atm^{-1}$ is the Henry's constant for $CO_2$ in water at $T = 298.15\ K$, i.e., $[CO_2] = (1.3 - 3.1) \times 10^{-3}\ M$.